# A Deep Reinforcement Learning Approach for Traffic Signal Control Optimization

Zhenning Li[a], Chengzhong Xu[a]*, Guohui Zhang[b]

[a] University of Macau
[b] University of Hawaii at Manoa
* Corresponding Author

# ABSTRACT

Inefficient traffic signal control methods may cause numerous problems, such as traffic congestion and waste of energy. Reinforcement learning (RL) is a trending data-driven approach for adaptive traffic signal control in complex urban traffic networks. Although the development of deep neural networks (DNN) further enhances its learning capability, there are still some challenges in applying deep RLs to transportation networks with multiple signalized intersections, including *non-stationarity environment*, *exploration-exploitation dilemma*, *multi-agent training schemes*, *continuous action spaces*, etc. In order to address these issues, this paper first proposes a multi-agent deep deterministic policy gradient (MADDPG) method by extending the actor-critic policy gradient algorithms. MADDPG has a centralized learning and decentralized execution paradigm in which critics use additional information to streamline the training process, while actors act on their own local observations. The model is evaluated via simulation on the Simulation of Urban MObility (SUMO) platform. Model comparison results show the efficiency of the proposed algorithm in controlling traffic lights.

**Keywords**: Multi-agent reinforcement learning; Deep learning; Adaptive traffic signal control; Transportation network

# INTRODUCTION

Traffic signal control is an efficient method of protecting traffic participants at intersections where multiple streams of traffic interact. Because of the capability of responding to fluctuating traffic demands, the adaptive signal control technology has been broadly implemented, and has attracted considerable interest in the research community (Wu *et al.*, 2017; Fu *et al.*, 2018; van de Weg *et al.*, 2018). By receiving and processing data from strategically placed sensors, adaptative control works based on real-time traffic dynamics about travel demand and traffic conditions and performance measurements of the control logic, such as traffic volume, queue length, travel speed and travel time (Wang, Huang and Lo, 2019). Compared to the fixed-time control, adaptive signal control has been proved to improve the quality of service that travelers experience on roadways, especially in busy urban areas, while reducing travel time by more than 10% on average (USDOT, 2017). However, inefficient adaptive control strategies can also cause numerous problems, such as forcing frustrated commuters to wait in congestion, resulting in more gasoline waste and air pollution, and even incurring vehicular accidents (Zhao *et al.*, 2018; Li *et al.*, 2019).

Numerous interdisciplinary approaches have been applied to improve the efficiency of adaptive signal control in the past two decades, including but not limited to fuzzy logic (Cheng *et al.*, 2016), case-based reasoning (De Schutter *et al.*, 2003; Elkosantini, Mnif and Chabchoub, 2011), artificial neural networks (Adeli and Samant, 2000; Ghanim and Abu-Lebdeh, 2015), genetic algorithm (Odeh *et al.*, 2015), and immune network algorithm (Darmoul *et al.*, 2017; Louati *et al.*, 2018). Among these technologies, reinforcement learning (RL), developed under the framework of Markov decision process (MDP), is a promising approach to learn adaptive signal control based on real-world traffic measurements. Different with conventional model-driven approaches or supervised learning algorithms, reinforcement learning is a type of goal-oriented algorithm, which learns how to achieve complex objectives (i.e., optimal control) in specific dimensions through many steps by learning from the experience interacting with complex transportation systems (Li, 2017).

As shown in Fig.1, in the RL framework for an isolated traffic signal control, the signal light is treated as an *agent*, and interacts with the traffic *environment* (i.e., traffic conditions) in a closed-loop MDP. The control *policy* is obtained by mapping from the traffic *states* (e.g., waiting time, queue length, total delay, etc.) to the corresponding optimal control *actions* (e.g., phase shift, cycle length change, green time increase, etc.). The agent iteratively receives a feedback reward for actions taken and adjusts the policy until it converges to the optimal control policy (El-Tantawy, Abdulhai and Abdelgawad, 2013). During the decision process, the policy that the agent takes combines both the exploitation of already learned policy and exploration of new policy that never met before. The RL algorithms can also be extended to a transportation network with multiple intersections. As the other settings remain the same, the goal of traffic signal agents changes to learn the optimal policies of each agent and to optimize the traffic condition of the whole traffic environment simultaneously. Studies using RLs to manipulate traffic signals are not rare in the last decade and have provided beneficial references for research (Abdulhai, Pringle and Karakoulas, 2003; Arel *et al.*, 2010; Prashanth and Bhatnagar, 2010).

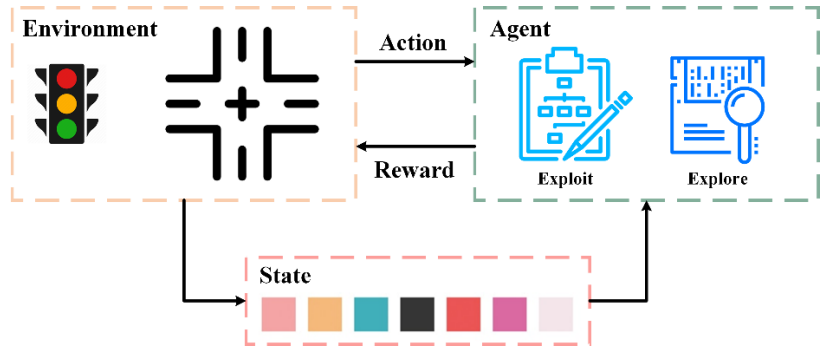

**FIGURE 1. Reinforcement learning framework for traffic light control**

With recent vast developments in the field of deep learning, we have been witnessing the renaissance of reinforcement learning, especially, the combination of deep neural networks (DNNs) and reinforcement learning, i.e., deep reinforcement learning (Schmidhuber, 2015). Deep RL has been prevailing in last several years with different applications, such as resources management in computer clusters (Mao *et al.*, 2016), games (Mnih *et al.*, 2013), natural language processing (NLP) (Young *et al.*, 2018), robotics (Kober, Bagnell and Peters, 2013), web system configuration (Bu, Rao and Xu, 2009), personalized recommendations (Wang *et al.*, 2014), etc. Deep RLs significantly outperform the traditional RLs since the deep learning architecture enables it to scale to decision-making problems that were previously intractable (Arulkumaran *et al.*, 2017). Considerable research efforts have been devoted to using deep RLs to solve operational problems in the domain of adaptive signal control in the past few years (Genders and Razavi, 2016; Li, Lv and Wang, 2016; Mousavi, Schukat and Howley, 2017). There are three major model-free deep RL approaches, including methods based on *value functions*, methods based on *policy search*, and methods which employ both value functions and policy search, namely *actor-critic* (AC) approach. More specifically, value function methods, for instance, *Q-learning* and *state-action-reward-state-action (SARSA)* algorithms, attempt to find a policy that maximizes the return by maintaining a set of estimates of expected returns for some policies (Zhao and Eskenazi, 2016). The state-action-value is updated by efficient bootstrapping, i.e., using current values of the estimate of the state-action-value to improve the estimate. These methods bear the flaws in practice due to the good convergence requires a stationary MDP transition, which is less likely in the domain of adaptive signal control. In addition, the policy degradation issue, i.e., current policy may still be far from optimal even if the value function estimation is already exceptionally accurate (Kakade and Langford, 2002). On contrast, policy search methods do not need to maintain a value function model, but directly search for an optimal policy. These methods assumed that the policy is stochastic and follows a parameterized distribution. This treatment enables these methods can learn stochastic policies and be effective in high-dimensional or continuous action spaces. However, the drawbacks are also significant, i.e., evaluating a policy is typically inefficient and high variance, and moreover, these methods always converge to a local rather than global optimum (Busoniu *et al.*, 2017). The actor-critic methods have grown in popularity as an effective means of combining value function with an explicit representation of the policy, and use the value function as a baseline for policy gradients. The fundamental difference between actor-critic methods and other baseline methods are that the former utilizes benefits of policy search methods with learned value function, which are able to learn from full returns and/or temporal difference (TD) errors (Schmidhuber, 2015). Recent papers have demonstrated that the actor-critic methods outperform other deep RLs in various applications (Lillicrap *et al.*, 2015; Wang *et al.*, 2016; Lowe *et al.*, 2017).

Although DNN significantly improves the applicability of RLs, there are still some challenges in applying deep RLs to transportation networks with multiple signalized intersections, including *non-stationarity environment*, *exploration-exploitation dilemma*, *multi-agent training schemes*, *continuous action spaces*, etc. The following contents will detailed describe these challenges. Firstly, in a single agent environment, an intersection only needs to concern the outcome of its individual actions. However, in the network domain, an intersection not only needs to observe the outcome of its own actions but also the behaviors of other intersections. In addition, all intersections are not homogeneous in real world since each intersection has different geometric characteristics, traffic components and volumes. The procedure is further complex because all intersections potentially need to interact with others and learn their actions concurrently. Therefore, this is a moving-target learning problem, i.e., learning among intersections may cause changes in the policy of an agent, and can affect the optimal policy of others. The estimated potential rewards on the action will be inaccurate, and thus a good strategy at a given point in the multi-agent setup will not be retained in the future, i.e., the optimal control policy of one intersection may not the networkwide optimal policy. At this condition, the traffic environment becomes non-stationary, and the Markov property no longer exists in this environment. This poses a huge challenge to the direct use of past experience replays, making Q-learning problematic to stabilize in such a multi-agent environment. On the other hand, policy gradient methods usually have a very high variance when multiple agents need to be coordinated (Lowe *et al.*, 2017).

In addition, collecting and processing information must be repeated to some extent, while ensuring that it does not affect the stability of the agent. Further complications of the exploration-exploitation dilemma arise due to the presence of multiple agents. The exploration-exploitation trade-off requires online RL algorithms that balance between the exploitation of the agent's current knowledge with exploratory information gathering actions taken to improve that knowledge (Buşoniu, Babuška and De Schutter, 2010). Agents have to not only explore information about the environment, but also the information of other agents in order to adapt to their behaviors. However, too much exploration can undermine the stability of other agents and affect the efficiency of the algorithm, making it more difficult to explore the agent's learning tasks.

Besides, most previous studies that directly extended the single agent deep RL to multi-agent environment is to learn each agent independently by considering other agents as part of the environment (Tampuu *et al.*, 2017). Due to structural limitations of traditional RLs, the method is prone to overfitting and is computationally expensive, and thus the number of agents involved is limited. Therefore, it is not surprising that those early attempts were only primarily aimed at controlling traffic on isolated intersections or small transportation networks with special attention on arterials (Bingham, 2001; Balaji,

German and Srinivasan, 2010). An alternative and popular approach is to use *centralized learning and decentralized execution*, where a group of agents can apply a centralized approach to training through open communication channels (Kraemer and Banerjee, 2016). The advantages of decentralized polices lie especially on partial observability and limited communication of agents. Centralized learning of decentralized strategies has become a standard paradigm for multi-agent setups because the learning process can be without communication constraints and additional state information can be obtained (Foerster *et al.*, 2018).

A recent paper introduced a multi-agent deep deterministic policy gradient (MADDPG) method based on the actor-critic policy gradient algorithms (Lowe *et al.*, 2017). MADDPG has a centralized learning and decentralized execution paradigm in which critics use additional information to streamline the training process, while actors act on their own local observations. This framework perfectly solves the networkwide adaptive traffic signal control problems which are difficult to be handled with other deep RLs, such as deep Q-learning (Zeng, Hu and Zhang, 2018), actor-critic (Wang *et al.*, 2016), etc. More specifically, the action value functions are learned in a centralized manner, implying that the actions of all intersections can be taken into account concurrently. Thus, the algorithm learns a collection of policies for each agent, rather than individual policies of a single agent. In each episode of the learning process, each intersection uniformly extracts a control policy from its ensemble. This treatment significantly reduces the non-stationarity caused by multiple agents learning simultaneously. In addition, at execution time the learned policies can only use local information (i.e., their own observations), so that the policies can be used by each agent without relying on further communication, which can significantly reduce the computational burden. In addition, previous deep RLs always assume the environment is discrete or differentiable, while the assumption is contrary to the real-world traffic environment dynamics. These advantages allow the MADDPG algorithm to formulate the adaptive traffic signal control problems in the traffic network environment consisting of both cooperative and competitive agents.

To the best of our knowledge, this is the first paper that formulates MADDPG for adaptive traffic signal control, by extending the algorithm of actor-critic policy gradient methods (specifically deep deterministic policy gradient, DDPG). In order to develop a robust and efficient control framework, several techniques are further proposed to deal with the issues arose by the multi-agent environment. We assume traffic conditions can be obtained by loop detectors, and different intersections can share their information. By this means, actions taken by all the others are known to each individual agent, thus the environment is stationary as the policies change. In addition, to reduce the computational burden and resolve the potential communication failure between agents, we relax the common assumption in other RLs that each agent's policies are always known to others. The remainder of the paper is organized as follows: In the Methodology Section, we first review the classical reinforcement learning algorithms, and then we carefully state our proposed MADDPG algorithm. In the Numerical Experiments Section, the proposed algorithm is evaluated by a real-world traffic network in Montgomery County, Maryland. Finally, the paper is concluded in the Conclusion Section.

# METHODOLOGY
## Problem Statement

We consider a road network with $N(N > 1)$ intersections, and each intersection has multiple lanes. Our goal is to find an optimal traffic control strategy with the aim of maximizing throughput in this network, i.e., the number of vehicles that have completed their trips in the road network during throughout a period. We formulated the problem in a reinforcement learning framework with four main components including: reward, state, action, and coordination strategy. A four-tuple $<\mathcal{S}, \mathcal{A}, \mathcal{R}, \mathcal{T}>$ is used to define the RL model, where $\mathcal{S}$ is the possible state space, $\mathcal{A}$ is the possible action space, $R$ denotes the reward space, and $\mathcal{T}$ is the transition function space among all states. An agent is an action executor that obtains a numerical reward by taking actions based on the environment and current state. The reward is the goal in a reinforcement learning problem. Equivalently, RL can be treated as an optimization method for the target which is specified as a reward function in the RL context. The agent takes an action according to the environment at each decision point, and the environment sends the reward back to the agent. In the long run, the agent's goal is to maximize the objective rewards through trial and error. A policy is built up of a series of consequent actions. The goal of reinforcement learning is to learn an optimal policy and maximize the cumulative expected rewards from the initial state.

In the network-wide traffic signal control problem, each signal light in the intersection is treated as an agent. The following contents give a description of actions, states, and rewards in this network. At each time step, the agent receives some quantitative information of the environment as the state representation. We assume that all intersections are equipped with near-intersection loop detectors which can detect real-time traffic conditions and send this information to the signal control center. Therefore, the traffic state data such as queue lengths, flow rate, vehicle speed, etc., can be easily obtained. In this study, we select two easily measurable elements including waiting time and volume as key indexes of the state of agents. The waiting time of a vehicle is defined as the time period a vehicle during which the vehicle is in a waiting status, i.e., the status that the vehicle with a speed of less than 0.1 m/s. The volume of a lane is defined as the number of vehicles on the lane, which equals to the sum of queueing vehicles and moving vehicles on the lane. More specifically, the state of intersection $i$ at time $t$ is defined as a two-value vector $s_{t,i} =< \sum w_{t,i}[l], \sum f_{t,i}[l] >, \forall l$, where $w_{t,i}[l]$ is the waiting time for the first vehicle in the each incoming lane $l$, and $f_{t,i}[l]$ is the flow rate of lane $l$.

Different action schemes have significant impacts on the performance of traffic signal control strategies. In current adaptive control algorithms, such as SCOOT and SCATS, there are several different control strategies, for instance, phase shift, phase length change, phase itself, etc., or a combination of some of the above. However, a complicated action design may considerably increase the computational burden. In this study, we follow the conventional adaptive control settings and define the action of each agent as phase switch. More specifically, the agent could choose one phase from the set of all feasible phases, in which the phase sequence is not predetermined. Consequently, this type of signal timing is more flexible, and the agent is learning to select the stage to change without assuming that the signal will change in a round-robin method.

Due to the complex spatial-temporal structure of traffic data, the reward should be spatially decomposable and can be easily measured after an action. Besides, since our primary goal is to maximize the total throughput of the network, the reward here is defined with two components, including the queue length and the sum of waiting time of vehicles on each individual incoming lane. In addition, different priorities could be assigned according to the traffic volume or emergency situations if needed. Let $L_p$ indicate the set of lanes on the roads that need to be assigned higher priority, then the reward of the intersection $i$ at time $t$ can be written as $r_{t,i} = -a_1 \left( \sum_{l \in L_{p,i}} a_2 q_{t+\Delta t,i}[l] + \sum_{l \notin L_{p,i}} b_2 q_{t+\Delta t,i}[l] \right) - b_1 \left( \sum_{l \in L_{p,i}} a_2 w_{t+\Delta t,i}[l] + \sum_{l \notin L_{p,i}} b_2 w_{t+\Delta t,i}[l] \right)$, where $q_{t+\Delta t,i}[l]$ is the sum of queue length over each approaching lane $l$, $a_1, a_2, b_1, b_2$ are non-negative parameters, $a_2 \geq b_2$, and $a_2 + b_2 = 1$. It is worth noting that since the reward is post-decision and thus both the two components are measured at time $t + \Delta t$ instead of at time $t$.

There are various algorithmic frameworks for RL methods from different perspectives. The RL method can be classified into a value-based method or a policy gradient method, respectively, depending on whether it is a learning value function or an explicit learning strategy parameter. The following content first reviews some classical reinforcement learning algorithms, and then proposes improved algorithms based on these algorithms to solve our problems.

**Background**

*A. Multi-agent Extension of Markov Decision Processes*

A multi-agent extension of Markov decision processes called partially observable Markov games is considered in this study (Littman, 1994). A Markov game for multiple $N$ ($N > 1$) agents is defined by a set of states $\mathcal{S}$, a set of actions, $\mathcal{A} = \{\mathcal{A}_1, \dots, \mathcal{A}_N\}$, and a set of observations $\mathcal{O} = \{\mathcal{O}_1, \dots, \mathcal{O}_N\}$, one for each agent on the environment. State transitions are controlled by the current state and one action from each agent: $\mathcal{T}: \mathcal{S} \times \mathcal{A}_1 \times \dots \times \mathcal{A}_N \mapsto PD(\mathcal{S})$. Each agent also has an associated reward function, $\mathcal{R}_i: \mathcal{S} \times \mathcal{A}_1 \times \dots \times \mathcal{A}_N \mapsto \mathfrak{R}$, and receives a private observation correlated with the state $\boldsymbol{o}_i: \mathcal{S} \mapsto \mathcal{O}_i$. The initial states are determined by a distribution $\rho: \mathcal{S} \mapsto [0,1]$. Each agent $i$ aims to maximize its expected sum of discounted rewards, $R_i = \sum_{t=0}^{T} \gamma^t r_i^t$, where $\gamma$ is a discount factor, $T$ is the time horizon, and $r_i^t$ is the reward received by agent $i$.

*B. Q-learning and Deep Q-networks*

Q-learning is a fundamental RL method that fits the Q-function with a parametric model $Q_\theta$. The action-value function in Q-learning for policy $\pi$ is given as $Q^\pi(s, a) = \mathbb{E}[R|s^t = s, a^t = a]$. This Q function can be recursively rewritten as $Q^\pi(s, a) = \mathbb{E}_{s'}[r(s, a) + \gamma \mathbb{E}_{a' \sim \pi}[Q^\pi(s', a')]]$. On the other hand, deep Q-networks (DQN) learns the action-value function $Q^*$ corresponding to the optimal policy by minimizing the loss:

$$\mathcal{L}(\theta) = \mathbb{E}_{s,a,r,s'}[(Q^*(s, a|\theta) - y)^2], \text{ where } y = r + \gamma \max_{a'} \bar{Q}^*(s', a') \qquad (1)$$

where $\bar{Q}$ is a target Q function whose parameters are intermittently updated with the latest $\theta$, which helps to stabilize learning.

The multi-agent extension of Q-learning can be achieved by having each agent $i$ learning an independently optimal function $Q_i$ (Tan, 1993). However, as aforementioned, since the agent independently updates its policies during the learning process, the environment seems to be non-stationary from the perspective of any agent, violating the Markov assumptions required for convergence of Q-learning. Another difficulty is that the experience replay buffer cannot be used in such a setting since in general, $P(s'|s, a, \pi_1, \ldots, \pi_N) \neq P(s'|s, a, \pi'_1, \ldots, \pi'_N)$ when any $\pi_i \neq \pi'_i$.

### C. Policy Gradient Algorithms

The major idea of policy gradient (PG) algorithms is to directly adjust the parameters $\theta$ of the policy in order to maximize the objective $J(\theta) = \mathbb{E}_{s \sim p^\pi, a \sim \pi_\theta}[R]$ by taking steps in the direction of $\nabla_\theta J(\theta)$. Using the Q function defined previously, the gradient of the policy can be written as:

$$\nabla_\theta J(\theta) = \mathbb{E}_{s \sim p^\pi, a \sim \pi_\theta}[\nabla_\theta \log \pi_\theta(a|s) Q^\pi(s, a)] \tag{2}$$

where $p^\pi$ is the state distribution. There are several methods to estimate $Q^\pi$, resulting in different practical algorithms, for instance, REINFORCE (Duan *et al.*, 2016), actor-critic (Konda and Tsitsiklis, 2000), etc. Specifically, the actor-critic algorithm uses an approximation, $Q_\varpi(s, a)$, to estimate the true action-value function $Q^\pi(s, a)$.

The multi-agent extension of PG algorithms also faces various challenges. In a multi-agent environment, since the reward of an individual agent is usually associated with the actions of other agents, the reward conditioned only on the agent's own actions (when the agent's optimization process does not consider the actions of other agents) exhibits much more variability, therefore increasing the variance of its gradients (Buşoniu, Babuška and De Schutter, 2010; Lowe *et al.*, 2017). Other algorithms that typically used to moderate high variance, for instance, actor-critic with baseline, are also problematic in multi-agent settings due to non-stationarity issues aforementioned (Foerster *et al.*, 2018).

### D. Deterministic Policy Gradient Algorithms

A recent extension of the policy gradient framework using deterministic policies $\mu_\theta: \mathcal{S} \mapsto \mathcal{A}$, named deterministic policy gradient (DPG), combined the benefits of DQN and actor-critic algorithms (Silver *et al.*, 2014). In particular, if the action space $\mathcal{A}$ and the policy $\mu$ are continuous (then the Q-function is presumed to be differentiable with respect to the action argument), the gradient of the objective $J(\theta) = \mathbb{E}_{s \sim p^\mu}[R(s, a)]$ can be rewritten as:

$$\nabla_\theta J(\theta) = \mathbb{E}_{s \sim \mathcal{D}}[\nabla_\theta \mu_\theta(a|s) \nabla_a Q^\mu(s, a)|_{a=\mu_\theta(s)}] \tag{3}$$

One of the major advantages of DPGs is that although stochastic policy gradients are integrated in both state and action spaces, DPGs are only integrated over the state space, requiring fewer samples in problems with large action spaces. Previous studies have shown that DPG greatly improves the stochastic policy gradient equivalence in high-dimensional continuous control problems (Silver *et al.*, 2014).

By integrating DNN, later work introduced Deep DPG (DDPG), which uses deep neural networks to concurrently approximate the policy $\mu$ and critic-function $Q^\mu$. It is also an off-policy algorithm, and can be thought of as being deep Q-learning for continuous action spaces.

**Multi-Agent Deep Deterministic Policy Gradient Algorithms (MADDPG)**

Considering the studied multi-intersection environment, each intersection is both cooperating and competing with other intersections simultaneously, resulting a mixed-cooperative-competitive environment. As aforementioned, the naïve policy gradient methods do not have outstanding performance in such a multi-agent environment, since the policies of each agent are unstable and the environment is non-stationary for each agent. That is, $P(s'|s, a, \pi_1, \dots, \pi_N) \neq P(s'|s, a, \pi'_1, \dots, \pi'_N)$ for any $\pi_i \neq \pi'_i$. To some extent, it does not make sense for agents to optimize their policies based on this unstable environment state. Policies that are optimized in the current state may not be effective in the next changed environment state. These problems are indeed caused by the fact that there is no interaction between the agents, and they do not know what policies the "teammates" or "opponents" will adopt, resulting in the selection of actions only according to their own circumstances, while ignoring the whole environment. Naturally, a promising way to solve this problem is to propose an algorithm that can take advantage of the information of other agents.

In the article of Lowe *et al.* (2017), the authors adopted a framework of *centralized training* with *decentralized execution*. The proposed algorithm can be regarded as a multi-agent extension of the actor-critic policy gradient methods where the critic is augmented with extra information about the policies of other agents. More specifically, in a game with $N$ agents with policies parametrized by $\boldsymbol{\theta} = \{\theta_1, \dots, \theta_N\}$, and let $\boldsymbol{\pi} = \{\boldsymbol{\pi}_1, \dots, \boldsymbol{\pi}_N\}$ be the set of all agent policies. The gradient of the expected return for agent $i$, $J(\theta_i) = \mathbb{E}[R_i]$, can be written as:

$$\nabla_{\theta_i} J(\theta_i) = \mathbb{E}_{s \sim p^\mu, a_i \sim \pi_i}\left[\nabla_{\theta_i} \log \boldsymbol{\pi}_i(a_i|o_i) Q_i^{\boldsymbol{\pi}}(\mathrm{x}, a_1, \dots, a_N)\right] \tag{4}$$

where $a_1, \dots, a_N$ are the $N$ agents, and $Q_i^{\boldsymbol{\pi}}(\mathrm{x}, a_1, \dots, a_N)$ is the *centralized action-value function* that takes as input of all these agents and state information x, and outputs the Q-value for agent $i$. It can be seen in Eq. (4), each $Q_i^{\boldsymbol{\pi}}$ is learned separately. Thus, agents can have indiscriminate reward structures, and even resulting in rewards conflict in competitive environments.

In order to let the above algorithm more accommodate to multi-agent settings, we can extend this idea by introducing deterministic policies. Let $\boldsymbol{\mu}_i$ be continuous policies with respect to parameters $\boldsymbol{\theta}_i$, then Eq. (4) can be written as

$$\nabla_{\theta_i} J(\mu_i) = \mathbb{E}_{\mathrm{x},a \sim \mathcal{D}}\left[\nabla_{\theta_i}\boldsymbol{\mu}_i(a_i|o_i) \nabla_{a_i} Q_i^{\boldsymbol{\mu}}(\mathrm{x}, a_1, \dots, a_N)|_{a_i = \mu_i(o_i)}\right] \tag{5}$$

where $\mathcal{D}$ indicates the experience reply buffer and contains the tuples $(\mathrm{x}, \mathrm{x}', a_1, \dots, a_N, r_1, \dots, r_N)$, recording experiences of all agents. The centralized action-value function $Q_i^{\boldsymbol{\mu}}$ is updated as:

$$\mathcal{L}(\theta_i) = \mathbb{E}_{\mathrm{x},a,r,\mathrm{x}'}\left[\left(Q_i^{\boldsymbol{\mu}}(\mathrm{x}, a_1, \dots, a_N) - y\right)^2\right] \tag{6a}$$

and
$$y = r_i + \gamma Q_i^{\mu'}(x', a'_1, ..., a'_N)|_{a'_j=\mu'_j(o_j)} \quad (6b)$$

where $\mu' = \{\mu_{\theta'_1}, ..., \mu_{\theta'_N}\}$ is the set of *target policies* with delayed parameters $\theta'_i$.

In order to obtain multi-agent policies that are more robust to changes in the policy of other agents in this mixed-cooperative-competitive environment, a collection of $K$ different sub-policies is trained. At each episode, one particular sub-policy is randomly selected for each agent to execute. Suppose that policy $\mu_i$ is an ensemble of $K$ different sub-policies with sub-policy $k$ denoted by $\mu_i^k$. For agent $i$, we are then maximizing the ensemble objective: $J_e(\mu_i) = \mathbb{E}_{k\sim\text{unif}(1,K), s\sim p^\mu, a\sim\mu_i^k}[R_i(s,a)]$. Since different sub-policies will be executed in different episodes, we maintain a replay buffer $\mathcal{D}_i^k$ for each sub-policy $\mu_i^k$ of agent $i$. Accordingly, we can derive the gradient of the ensemble objective with respective to $\theta_i^k$ as follows:

$$\nabla_{\theta_i^{(k)}} J_e(\mu_i) = \frac{1}{K}\mathbb{E}_{x,a\sim\mathcal{D}_i^{(k)}}\left[\nabla_{\theta_i^{(k)}}\mu_i^k(a_i|o_i)\nabla_{a_i}Q^{\mu_i}(x, a_1, ..., a_N)|_{a_i=\mu_i^k(o_i)}\right] \quad (7)$$

The detailed learning schemes and algorithm procedure are shown in Fig. 2 and Algorithm 1, respectively.

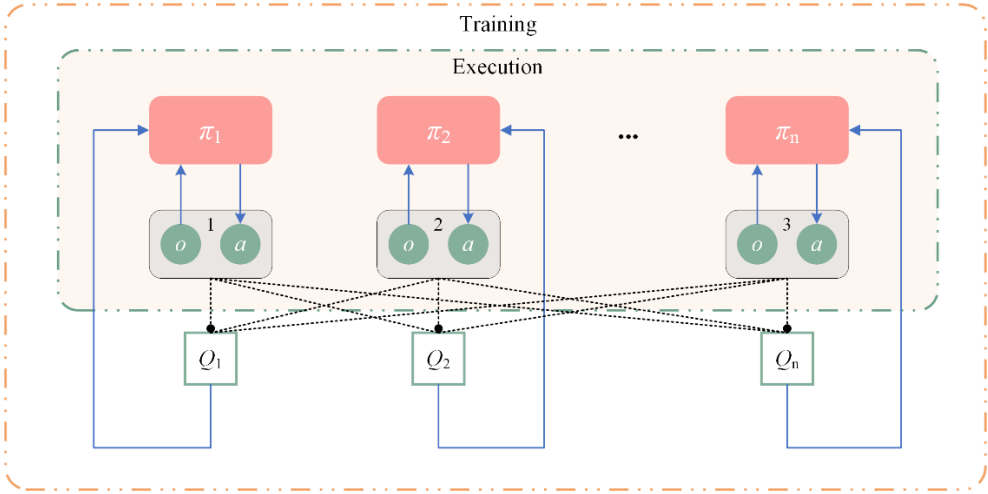

**FIGURE 2. Centralized Learning and Decentralized Execution Structure of MADDPG**

**Algorithm 1** Multi-Agent Deep Deterministic Policy Gradient for $N$ agents

```
1   for episode = 1 to M do
2       Initialize a random process 𝒩 for action exploration
3       Receive initial state x
4       for t = 1 to max-episode-length do
5           for each agent i, select action a_i = μ_{θ_i}(o_i) + 𝒩_t w/r/t the current policy and exploration
6           Execute actions a = (a_1, ..., a_N) and observe reward r and new state x'
7           Store (x, a, r, x') in replay buffer 𝒟
8           x ← x'
9           for agent i = 1 to N do
10              Sample a random minibatch of S samples (x^j, a^j, r^j, x'^j) from 𝒟
11              Set y^j = r_i^j + γ Q_i^{μ'}(x'^j, a'_1, ..., a'_N)|_{a'_k = μ'_k(o_k^j)}
12              Update critic by minimizing the loss ℒ(θ_i) = (1/s) Σ_j (y^j − Q_i^μ(x^j, a_1^j, ..., a_N^j))^2
13              Update actor using the sampled policy gradient:
```
$$\nabla_{\theta_i} J \approx \frac{1}{S} \sum_j \nabla_{\theta_i} \mu_i(o_i^j) \nabla_{a_i} Q_i^\mu(x^j, a_1^j, \ldots a_i, \ldots, a_N^j)\big|_{a_i = \mu_i(o_i^j)}$$
```
14          end for
15          Update target network parameters for each agent i:
16                                  θ'_i ← τθ_i + (1 − τ)θ'_i
17      end for
18  end for
```

## NUMERICAL EXPERIMENTS

### General Setups

The proposed MADDPG algorithm is evaluated in the Simulation of Urban MObility (SUMO) simulation platform, using a real-world complex traffic network extracted from Montgomery County, Maryland (as shown in Fig. 3). SUMO is an open-source traffic simulation package designed by the Institute of Transportation Systems at the German Aerospace Center to handle large road networks (Krajzewicz *et al.*, 2012). The multi-agent extension of the OpenAI Gym framework is utilized to setup the simulated environment (Brockman *et al.*, 2016). We used Python APIs provided by SUMO to convert the network from Open Street Map, and detailed intersection structures are shown in Fig. 4. In order to test the robustness and effectiveness of the proposed MADDPG algorithm, the traffic demand data consisting both PM peak hours (17:30-18:00) and congestion recovery hours (18:00-18:30) are used as the input to simulate the recurrent congestion situation. The daily Origin-Destination (OD) matrices are extracted from the Montgomery-National Capital Park and Planning Commission, and the traffic volumes are obtained from Maryland Department of Transportation and utilized to estimate the OD matrix from the raw daily OD matrices. For the unsignalized intersections, the default control logic in SUMO are utilized. In the simulation setting, the maximal speed of a vehicle is 60 mile/h, and the maximal acceleration and deceleration rates are 3.3 feet/s$^2$ and 15 feet/s$^2$, respectively. The duration of yellow signals is set to be 4 seconds.

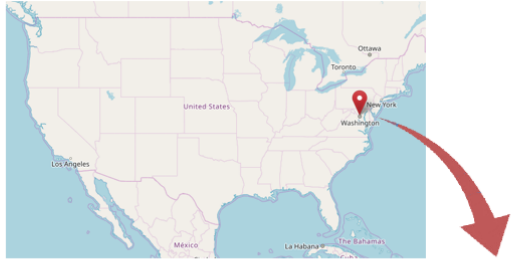

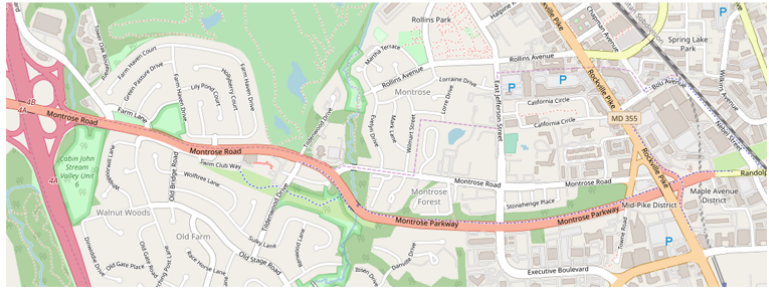

**FIGURE 3. Traffic Network in Montgomery County**

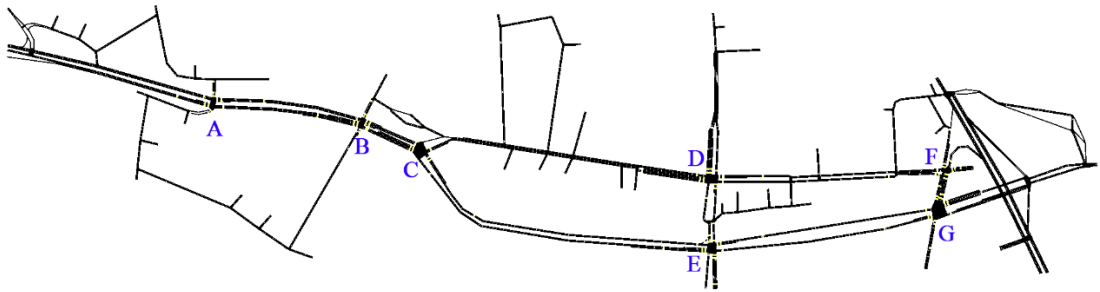

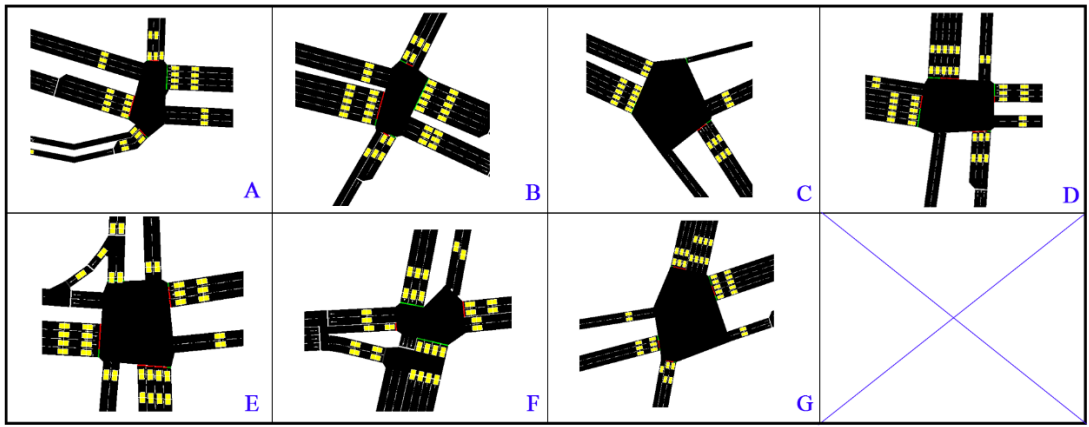

**FIGURE 4. Detailed Traffic Network Structure**

The MADDPG algorithm is trained in episodes, and different random seeds are used for generating different training and evaluation episode. The length of training and control step is one second, and thus, one episode contains totally 3,600 training steps. As shown in Fig. 5, the deep neural network architecture of our proposed MADDPG consists two networks, i.e., the actor network $\pi$ and the critic network $Q$. The actor network receives the current transportation state obtained by loop detectors and outputs the actions. More specifically, it starts with several fully connected layers with non-linear Leaky Rectified Linear Unit (ReLU) activations (Maas, Hannun and Ng, 2013). Thus, the width of the network is the same as the traffic light phases. A batch normalization layer and another fully connected layer with ReLU activation is designed following the last-mentioned layer to scale the output into a reasonable range. In addition, noise is injected as another layer prior to phase adjustment to keep exploration. The critic network receives the current state of simulation and actions generated by all agents, and outputs the Q-values associated to them. Like the actor, it is comprised of several fully connected layers with leaky ReLU activations, and a final fully connected layer with linear activation. Detailed hyperparameters of the MADDPG algorithm are presented in Table 1.

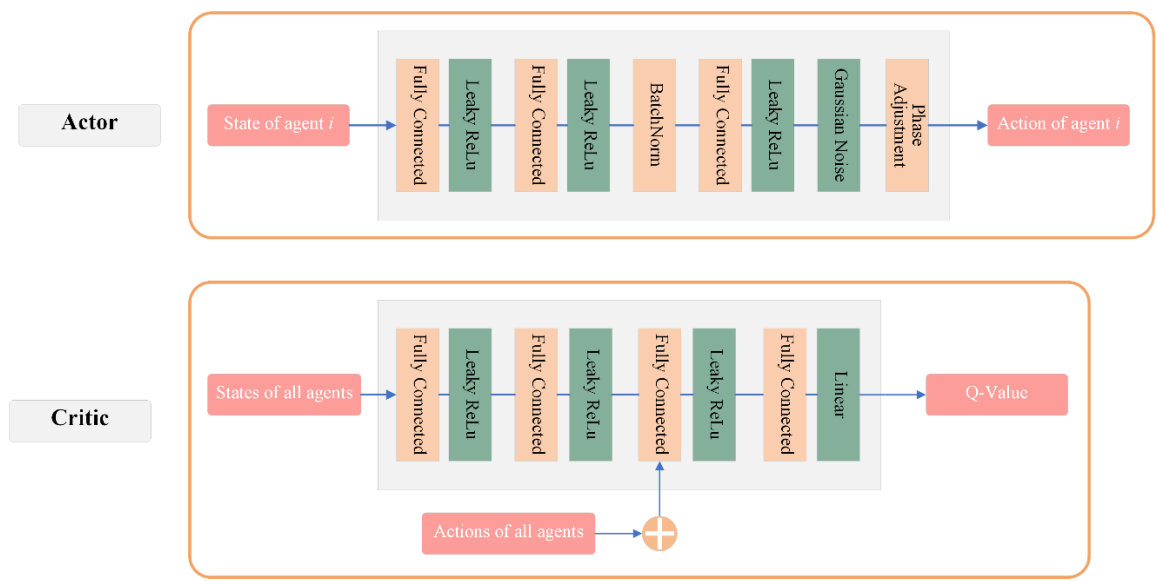

**FIGURE 5. Network Structure of MADDPG**

**TABLE 1. Values of Selected Hyperparameters**

| Parameter | Value | Description |
|---|---|---|
| $M$ | 100000 | Replay memory buffer size |
| $B$ | 512 | Minibatch size |
| $\gamma$ | 0.95 | Discount factor |
| $\tau$ | 0.01 | Soft update of target parameters |
| $\varepsilon_a$ | 0.001 | Actor learning rate |
| $\varepsilon_c$ | 0.001 | Critic learning rate |
| $w$ | 0 | L2 weight decay |
| $r_c$ | 1.0 | Reward scaling (1.0 means no scaling) |
| $\sigma$ | 0.01 | Noise standard deviation |
| $l$ | 1 | Frequency of performing learning step |
| $g_c$ | True | Gradient clipping true or false |
| $g_v$ | 1 | Gradient clipping value |
| $FC_1$ | 64 | Input channels for the first hidden layer |
| $FC_2$ | 64 | Input channels for the second hidden layer |

The performance of the proposed algorithm is compared with the fixed time traffic controls and two well-trained state-of-the-art benchmark algorithms in a simulation replication with the same random state, including DQN (Li, Lv and Wang, 2016) and DDPG (Casas, 2017). In the DQN implementation, a centralized control method is designed, i.e., an agent manages all intersections simultaneously. On the other hand, the network architectures used in DDPG is the same as those in MADDPG. To ensure fair comparison, all controllers have the same action space, state space, and interaction frequency. In addition, different hyperparameters for benchmarks are tested for several times to ensure adequate performance, and fine-tuned values are injected in the two models. Finally, all algorithms are trained with 1,000 episodes.

**Experiment Results**

As shown in Fig. 6(a), the average reward of all episodes for MADDPG, DQN, and DDPG algorithms are -48.0, -79.5, and -152.9, respectively. On the other hand, the maximum reward for these three algorithms are -14.0, -26.9, and -39.73, respectively. In addition, MADDPG remains remarkably stable once it reached its peak performance, while the other two are unstable and difficult to converge. Therefore, it is clearly that MADDPG outperforms others for this given objective. The reason is that Q-learning is difficult to maintain stable in such a non-stationary environment, while DDPG is challenged by the high variance when multiple intersections need to be coordinated simultaneously.

The average queue lengths and delays per vehicle networkwide for different algorithms are shown in Figs. 6(b) and 6(c). All these comparison results are extracted from the best preformed episode of these algorithms. The average queue lengths for MADDPG, DQN, DDPG, and fixed time controllers are 0.8 veh, 2.3 veh, 5.0 veh, and 9.8 veh, respectively. The average delays for the four controllers are 5.2 s/veh, 9.06 s/veh, 13.4 s/veh, and 21.2 s/veh, respectively. It should be noted that even though the traffic volumes and ODs are the same for all these controllers, the trends of queue lengths and delays are different because the controllers generated different control strategies. These results show that all reinforcement learning algorithms have better control performance than the fixed time control, and have the abilities to recover from congestion. As contrast, the fixed time control has the worst performance, and the queues are difficult to mitigate even though the traffic volume becomes lower at the last 30 minutes. Among all these deep RL algorithms, MADDPG provides the most stable control policy since it has the smallest variance, and has the shortest average queue length and smallest average delay.

The throughputs for different algorithms are presented in Fig. 6(d). The flow rate is aggregated into one minute to provide significant comparison results. It can be seen that at the first beginning, there are not many differences of traffic throughput between the four controllers. However, in the middle part of the simulation (20-40min), MADDPG has the largest output flow beating all others, while fixed time control has the smallest flow rate and thus fails to quickly eliminate congestions. In the last part of the simulation, MADDPG still stays ahead but the advantage is not particularly obvious. Totally, in this one-hour simulation, 6118, 5596, 5511, and 4977 vehicles finished their trips under the control of MADDPG, DQN, DDPG, and fixed time, respectively. These results also evidenced that our proposed model has better performance than other peers.

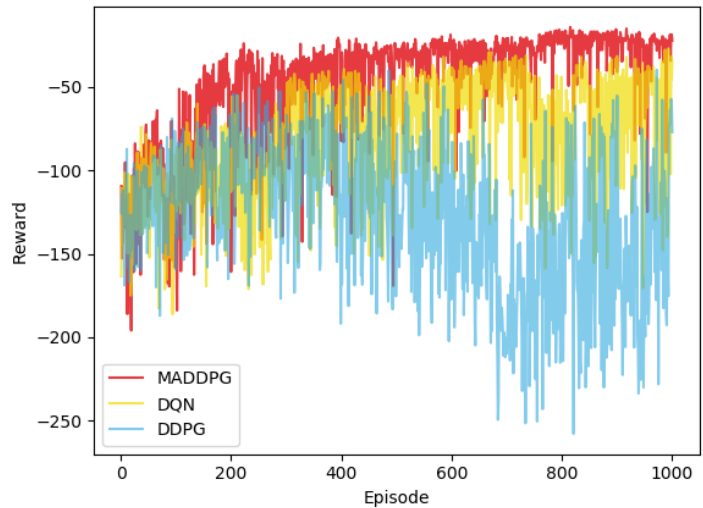

**FIGURE 6(a). Training Curves of Different Algorithms**

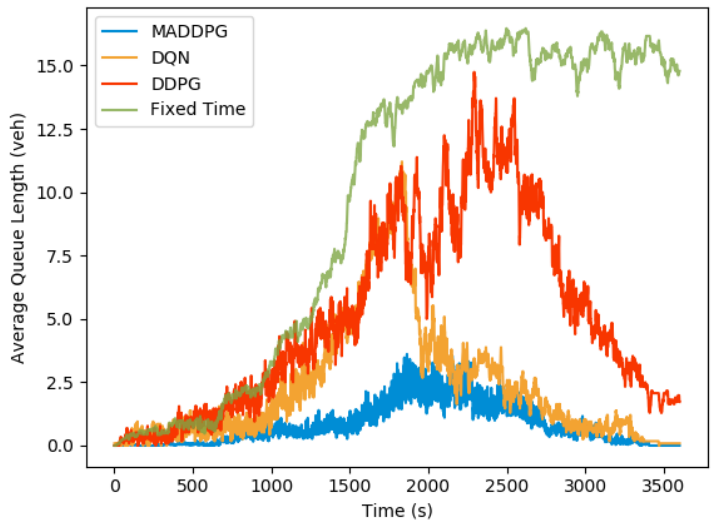

**FIGURE 6(b). Average Queue Length Comparison of Different Algorithms**

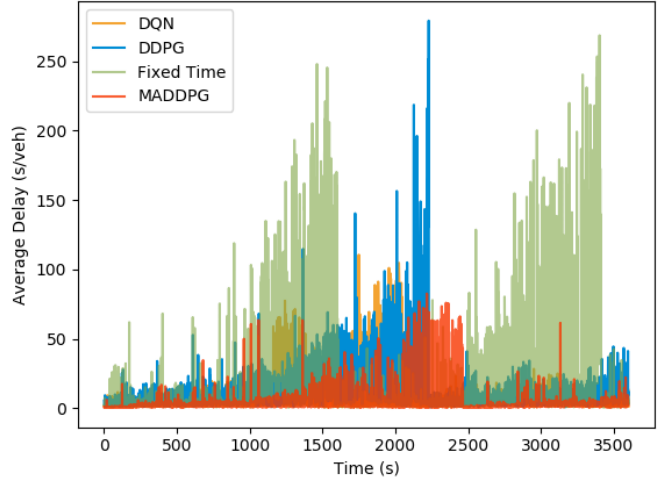

**FIGURE 6(c). Average Delay Comparison of Different Algorithms**

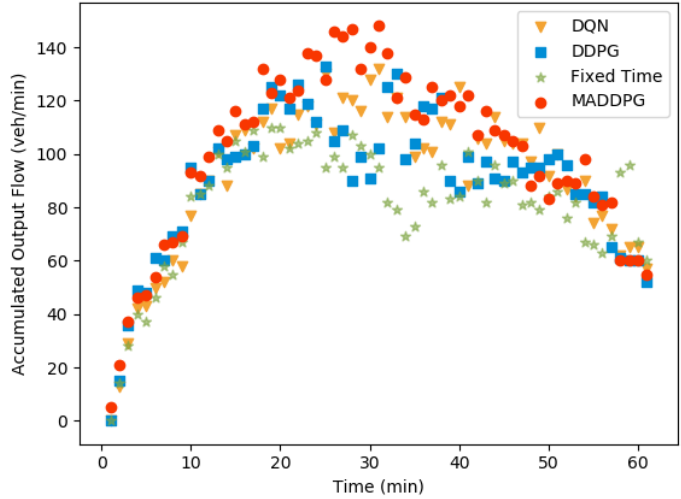

**FIGURE 6(d). Accumulated Output Flow Comparison of Different Algorithms**

**CONCLUSIONS AND LIMITATIONS**

In this paper, a multi-agent deep deterministic policy gradient algorithm is proposed to formulate the adaptive traffic signal control problems. The proposed algorithm is evaluated in the Simulation of Urban MObility (SUMO) simulation platform, using a real-world complex traffic network extracted from Montgomery County, Maryland. Experiment results demonstrate that the robustness, optimality, and scalability of the proposed MADDPG algorithm, which outperformed other state-of-the-art benchmark algorithms.

There are still some limitations for real-world implementation of the proposed algorithm. First, different definitions in formulating the reward functions and state characteristics could significantly affect the control outcomes. Furthermore, the state representation can be in a high dimensional space, especially when dynamic traffic conditions are used as part of the state representation. In this case, the neural network will require more training data samples and become more difficult to converge. Therefore, how to learn efficiently (e.g., learning from limited data samples, efficient exploration-exploitation, etc.) is a critical question in the application of RL in traffic signal control.

**ACKNOWLEDGEMENT**